# How the SiC substrate impacts graphene atomic and electronic structures

**L. Magaud[*,1], F. Hiebel[1], F. Varchon[1], P. Mallet[1], J.-Y. Veuillen[1]**

[1] Institut Néel, CNRS and UJF, BP 166, 38042 Grenoble cedex 9, FRANCE

[*] Corresponding author: e-mail laurence.magaud@grenoble.cnrs.fr

Graphene, the two-dimensional form of carbon presents outstanding electronic and transport properties. This gives hope for the development of applications in nanoelectronics. However, for industrial purpose, graphene has to be supported by a substrate. We focus here on the graphene-on-SiC system to discuss how the SiC substrate interacts with the graphene layer and to show the effect of the interface on graphene atomic and electronic structures.

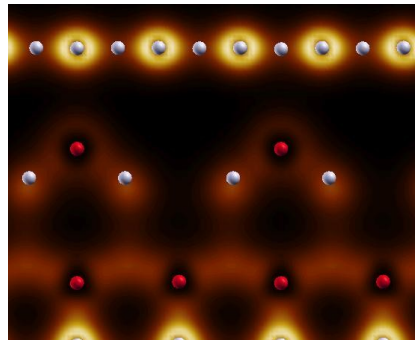

Figure : Cross section of the ab initio total charge density at the graphene / C-terminated SiC interface.

**Introduction.** Because of its exceptional electronic properties [1,2,3,4], graphene has recently attracted a lot of attention. Graphene is the two-dimensional (2D) form of carbon; it consists in one isolated plane of C atoms arranged on a honeycomb lattice. The unit cell contains two C atoms that generate two inequivalent sublattices. All graphene outstanding properties originate in this honeycomb lattice from which two main electronic properties derive : i) a linear dispersion of the bands in the vicinity of the Fermi level [5] and ii) a special symmetry of the wavefunction, the so-called chirality [3,6]. A famous manifestation of this peculiar electronic structure is the anomalous quantum Hall effect that has been observed up to room temperature [7]. i) and ii) are the fingerprints of an ideal graphene layer, how they evolve when it lies on a substrate is a hot question since this is indeed required for applications. We focus here on the graphene-on-SiC system where the annealing of a SiC substrate leads to the formation of few graphitic layers [2,8]. Though this system consists in several C layers on top of a SiC substrate, transport measurements, Raman and Landau Level spectroscopies evidence properties expected for an isolated graphene sheet [9-11]. The question then is how the conducting graphene layer(s) can be decoupled from its neighborhood (substrate and other C layers). On the basis of extensive ab initio calculations and STM experiments, we consider the interaction of the first C layer with the SiC surface and compare the results for the two polar surface terminations, (0001) (Si terminated) or (000-1) (C terminated), of an hexagonal SiC polytype.

**Methods.** Calculations are performed with the code VASP that is based on the density functional theory [12]. We use the Perdew and Wang formulation of the generalized gradient approximation [13]. The interface is modeled by a slab and dangling bonds on the other surface are saturated by H atoms. For all calculations, the ultrasoft pseudopotentials, the k-points sampling and the size of the supercell have been extensively tested. The plane wave basis cutoff is equal to 211 eV. Residual forces are lower than 0.02 eV/Å for converged structures. When graphene is involved, the K point of the graphene 1x1 Brillouin zone has to be included in the k-point mesh to get a correct description of the Fermi level.

Graphene samples are grown in a UHV chamber. Cleaning is first made by annealing a 6H SiC substrate for 20 mn under a Si flux at 850°C. Further successive annealings at higher temperature allows a controlled growth of graphene (graphitization occurs at 1350°C on the Si face and 1150°C on the C face). The samples are characterized with usual surface science techniques (LEED, Auger) prior to transfer in our STM chamber.

**Graphene on the (0001) (Si) face of SiC.**
The Si surface of hexagonal SiC polytypes goes through different reconstructions when the temperature is increased : 3x3 at 850°C under Si flux, √3x√3 R30 below 1000°C, 6√3x6√3R30 (6R3) at 1150°C and graphitization at 1350°C. The 6R3 phase was first assumed to be due to multiple diffusion but it is now established that the reconstruction comes from a C honeycomb lattice distorted by a strong interaction with the substrate [11,14-17]. Interaction with the substrate was modeled in a very large supercell -more than 1200 atoms- that describes the actual experimental interface structure [14,17,18]. Covalent bonds are formed between the Si atoms of the last SiC layer and the first C layer. This prevents any graphitic electronic properties for this layer. The graphitic nature of the film is recovered by the second and the third absorbed C layers. They are electron doped by a charge transfer from the substrate in agreement with ARPES experiments [5,19]. The complex geometry of the first carbon layer generates soft ripples in the honeycomb lattice of the graphene - second C- layer. The corrugation induced by the substrate extends over the two first C layers. On the other hand, LEED and STM [2,17,18] experiments demonstrate that the graphene is epitaxial on this face. The long range orientation of the graphene planes is then imposed by the substrate.

**Graphene on the (000-1) (C) face of SiC**

On the C face, STM images show the coexistence of two reconstructions (3x3 and 2x2) of the SiC surface that remain even when covered by one C layer [20]. No atomic model is established for the 3x3 reconstruction. We have recently proposed a model for the other interface [21], based on the SiC 2x2 native reconstruction [22]. In the bulk truncated 2x2 surface, we have 4 dangling bonds (DB) per unit cell. A Si adatom saturates 3 DB and one C atom (rest atom) remains unbounded. Furthermore, a charge transfer occurs from the adatom to the restatom so that the SiC surface is passivated. It is then semiconducting with either filled or empty DB states in the vicinity of the Fermi level. Interaction between graphene and SiC is then much smaller than on the Si face. In our calculation, a 5x5 graphene cell is superimposed to a 4x4 SiC cell (Figure 1a) – these two cells are nearly commensurate-. The total energy remains nearly unchanged when the graphene – SiC surface relative position is shifted. This is consistent with the rotational disorder observed experimentally in the

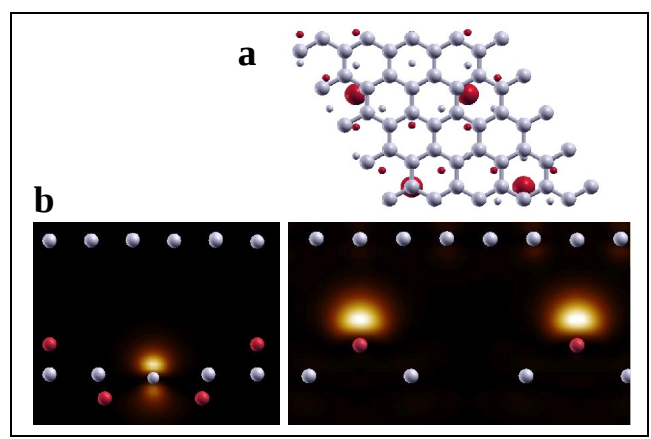

**Figure 1** (a) 4x4 (SiC) – 5x5(graphene) interface cell. C is in grey (light grey) and Si in red (dark grey). The size of the atomic spheres is proportional to the distance to the interface. b) and c) cross section salong z of |Ψ|² integrated from -0.75 to 0 eV (b) and 0. to 0.75 eV ( c). It shows the extension of the interface states.

graphene layers [19, 20, 23, 24]. This rotation is very important since it leads to an effective decoupling of adjacent C layers and a linear free graphene like dispersion [24]. A small interaction is observed between the SiC 2x2 surface and the C layer but at variance with the Si face, here, the band structure shows a linear dispersion characteristic of graphene from the first C layer in agreement with ARPES [19]. Figure 1 shows the extension of the states related to the SiC 2x2 surface. The square modulus of the wavefunction |Ψ|² is integrated from -0.75 to 0. eV (Fig. 1b) or from 0. to 0.75 eV (Fig. 1c), this corresponds to the energy range where the graphene dispersion and the filled or empty DB states coexist. No bonds can be seen, only a small intensity is observed for empty states between the adatom and the C atom that are located on top of each other.

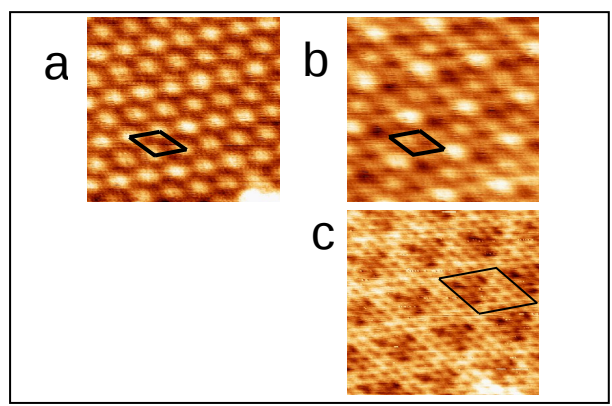

**Figure 2:** 4x4nm² STM images of graphene on 2x2-SiC (0° désorientation i.e. the computed configuration). (a), (b) Dual bias images, sample bias: -1.5V (a) (+1.5V (b)), It=0.2nA. A 2x2-SiC cell is shown on both images. (c) Sample bias: -50mV, It=0.2nA. A 4x4-SiC (5x5-G) cell is indicated.

Experimentally, high bias STM images only show the underlying 2x2-SiC structure [20] with C restatoms (Si adatoms) in filled (empty) states images, presenting the expected shift between them (figure 2 (a), (b)). The 4x4-SiC superstructure clearly shows up on the Si adatoms (figure 2 (b)). Low bias images (figure 2 (c)) reveal a graphene honeycomb structure with small perturbations leading to a 5x5-G (4x4-SiC) superstructure, consistently with the ab initio results.

**Conclusion.** Extensive ab initio calculations show that the morphology and the electronic properties of the C layers strongly depend on the SiC face used for the growth. These results are in very good agreement with our STM experiments and explain, for the C face, how the graphene layers can exhibit isolated graphene properties. On the Si face, the C layers are in epitaxy on SiC. The substrate surface is passivated by the first C layer – the bufferlayer- and interface extends on 2 C layers. On the C face, the native 2x2 reconstruction saturates the DB states so that the first C layer can already exhibit graphene properties. The counterpart of this small interaction is that the long range orientation of the graphene layer is not imposed by the substrate. A balance has to be found between a long range order in the graphene layer that can be imposed by the substrate (strong interaction – Si face) and the preservation of graphene electronic structure (small interaction – C face). Decoupling the graphene layers from the substrate after the growth could reveal to be interesting.

**Acknowledgements** Part of this work was supported by computer grants at IDRIS (CNRS) and CIMENT (phynum project). Fundings was provided by the ANR GRAPHSIC and the CIBLE Region Rhône Alpes projects.

### References

[1] A. H. Castro Neto, F. Guinéa, N. M. R. Peres, K. S. Novoselov and A. K. Geim, Rev. Mod. Phys. 81, 109 (2009).
[2] C.Berger et al, J.Phys. Chem. B108, 19912 (2004).
[3] K.S.Novoselov et al, Nature (London) 438, 197 (2005).
[4] Y.Zhang, Y.-W. Tan, H.L.Stormer, P.Kim, Nature (London) 438, 201 (2005)
[5] A.Bostwick, T.Ohta, Th.Seyller, K.Horn, E.Rotenberg, Nat.Pays. 3, 36 (2007).
[6] I.Brihuega, P.Mallet, C.Bena, S.Bose, C.Michaelis, L.Vitali, F.Varchon, L.Magaud, K.Kern, J.-Y.Veuillen, Phys. Rev. Lett 101, 206802 (2008).
[7] K.S.Novoselov et al, Science 315, 1379 (2007)
[8] I.Forbeaux, J.-M.Themlin, J.-M.Debever, Surf.Sci. 442, 9 (1999).
[9] C.Berger et al, Science 312, 1191 (2008).
[10] M.L.Sadowski, G.Martinez, M.Potemski, C.Berger, W.A. De Heer, Phys. Rev. Lett. 97, 266405 (2006)
[11] F.Varchon, R.Feng, J.Hass, X.Li, B.Ngoc Nguyen, C.Naud, P.Mallet, J.-Y.Veuillen, C.Berger, E.H.Conrad, L.Magaud Phys. Rev. Lett. 99, 126805 (2007)
[12] G.Kresse and J.Hafner, Phys. Rev. B 47, 558 (1993)
[13] J.P.Perdew and Y.Wang, Phys. Rev. B 33, 8800 (1986).
[14] F.Varchon, P.Mallet, J.-Y.Veuillen, L.Magaud, Phys. Rev. B77, 235412 (2008).100, 176802 (2008).
[15] A.Mattausch, O.Pankratov, Phys. Rev. Lett. 99, 076802 (2007).
[16] S.Kim, J.Ihm, H.J.Choi, Y.-W.Son, Phys. Rev. Lett.
[17] C.Riedl, U.Starke, J.Bernhardt, M.Franke, K.Heinz, Phys. Rev. B 76, 245406 (2008).
[18] P.Mallet, F.Varchon, C.Naud, L.Magaud, C.Berger, J.-Y.Veuillen, Phys. Rev.B76, 041403R (2007).
[19] K.V.Emtsev, F.Speck, Th.Seyller, L.Ley, J.D.Riley, Phys. Rev.B 77, 155303 (2008).
[20] F.Hiebel, P.Mallet, F.Varchon, L.Magaud, J.-Y.Veuillen, Phys. Rev. B 78, 153412 (2008).
[21] L.Magaud, F.Hiebel, F.Varchon, P.Mallet, J.-Y.Veuillen, Phys. Rev. B 79, 161405 (2009).
[22] A. Seubert, J.Bernhardt, M.Nerding, U.Starke, K.Heinz, Surf. Sci. 454-456, 45 (2000).
[23] F.Varchon, P.Mallet, L.Magaud, J.-Y.Veuillen, Phys. Rev. B 77, 165415 (2008).
[24] J.Hass, F.Varchon, J.E.Millan-Otoya, M.Sprinkle, N.Sharma, W.A.de Heer, C.Berger, P.N.first, L.Magaud, E.H.Conrad, Phys. Rev.Lett 100, 125504 (2008).